\documentclass{article}
\usepackage{graphicx}
\usepackage[centertags]{amsmath}
\usepackage{amsfonts}
\usepackage{amssymb}
\usepackage{amsthm}

\title{QUANTIZATION OF SYSTEMS WITH INTERNAL DEGREES OF FREEDOM IN TWO-DIMENSIONAL MANIFOLDS}

\author{Ewa Eliza Ro\.zko*,\\ Ewa Gobcewicz**\\
Institute of Fundamental Technological Research,\\
Polish Academy of Sciences,\\
$5^{\rm B}$, Pawi\'{n}skiego str., 02-106 Warsaw, Poland\\
e-mail: *erozko@ippt.pan.pl, **gobicz@op.pl}

\begin{document}

\maketitle
  
\begin{abstract}
Presented is a primary step towards quantization of infinitesimal
rigid body moving in a two-dimensional manifold. The special
stress is laid on spaces of constant curvature like the
two-dimensional sphere and pseudosphere (Lobatschevski space).
Also two-dimensional torus is briefly discussed as an interesting
algebraic manifold.
\end{abstract}


\section{General formulation}

It is well known that in a general Riemann space there is no
concept of extended rigid body, because, as a rule, the isometry
group is trivial. In constant-curvature spaces there are isometry
groups of maximal dimension $n(n+1)/2$, where $n$, obviously, denotes
the manifold dimension. However, even in this case literally
followed rigid body mechanics in flat spaces would be rather
doubtful. In the Riemann space there exists, however, a
well-defined concept of infinitesimal rigid body. The
configuration space of infinitesimal gyroscope in a Riemann space
$(M,g)$ is identified with $F(M,g)$ --- the principal fibre bundle of
orthonormal frames. Its dimension equals, of course $n(n+1)/2$;
there are n translational degrees of freedom and $n(n-1)/2$ internal
ones. Coordinates $x^{i}$ in $M$ induce in a natural way
coordinates $\left(x^{i},e^{i}_{A}\right)$ on $FM$ --- the bundle of any linear
frames in $M$. The quantities $e^{i}{}_{A}$ are components of the
moving frame $e_{A}$, $A=1,\ldots,n$ describing internal degrees of
freedom. If the internal motion is $g$-rigid, then, by definition,
the quantities $e^{i}{}_{A}$ are confined by constraints
\begin{equation}\label{eq.1}
g_{ij}\,e^{i}{}_{A}\,e^{j}{}_{B}=\delta_{AB}.
\end{equation}

Being restricted by these conditions, the quantities $e^{i}{}_{A}$ cannot be used as generalized coordinates. To calculate anything explicitly, one must have, however, some non-redundant configuration variables. The best way to achieve this
is to introduce some auxiliary non-holonomic reference frame
\cite{Gol01, Gol02, Gol03, Gol04S, Sl_2004, Prace IPPT_1, Prace IPPT_2, ROMP_SG, ROZ_2010, ROZ_2004} some fixed field of orthonormal bases $E_{A}$ defined all over the manifold $M$. Then, when at the time instant t the spatial
position is given by $x^{i}(t)$, and the internal parameters by
$e^{i}{}_{A}(t)$ ($e_{A}(t)$ are vectors attached at $x(t)\in M$), we have
\begin{equation}
e_{A}(t)=E_{B}(x(t))\,R^{B}{}_{A}(t),
\end{equation}
where $R\in {\rm SO}(n,\mathbb{R})$ is a proper orthogonal matrix \cite{God03, SO4},
\begin{equation}\label{eq.2}
\delta_{AB}=\delta_{CD}\,R^{C}{}_{A}\,R^{D}{}_{B}.
\end{equation}

Orthogonal groups ${\rm SO}(n,\mathbb{R})$ are parameterized in a variety
of standard ways, like Euler angles, rotation vector, etc. In this
way the configuration space $F(M,g)$ becomes explicitly
parameterized. Usually the choice of $E_{A}$ must be somehow
adapted to geometry of $(M,g)$ if calculations are to be
effective. The co-moving components of angular velocity
$\Omega^{A}{}_{B}$ are defined by the following formula:
\begin{equation}\label{eq.3}
\frac{De_{A}}{Dt}=e_{B}\,\Omega^{B}{}_{A},
\end{equation}
where obviously $D/Dt$ denotes the $g$-covariant
differentiation along the curve representing the translational
motion in $M$. The quantity $\Omega$ is skew-symmetric,
\begin{equation}\label{eq.4}
\Omega^{A}{}_{B}=-\delta_{BC}\,\delta^{AD}\Omega^{C}{}_{D}.
\end{equation}

The total kinetic energy (translational+internal)is given
analytically by
\begin{equation}\label{eq.5}
T=\frac{M}{2}\,g_{ij}\frac{dx^{i}}{dt}\frac{dx^{j}}{dt}+
\frac{1}{2}\,\delta_{AB}\,\Omega^{A}{}_{C}\Omega^{B}{}_{D}J^{CD},
\end{equation}
where $J$ is the inertial tensor in co-moving representation. This
quantity is, as usual, equivalent to some metric tensor $G$ on the
manifold $F(M,g)$, and
\begin{equation}\label{eq.6}
T=\frac{1}{2}\,G_{ab}\frac{dq^{a}}{dt}\frac{dq^{b}}{dt},
\end{equation}
where  $q^{a}$, $a=1,\ldots,n(n+1)/2$ is the total system of
generalized coordinates introduced by the choice of holonomic
coordinates $x^{i}$ and non-holonomic frame $E_{A}$ in $M$.

The simplest way of quantization is the Schr\"{o}dinger procedure \cite{Weyl}.
Every Riemann manifold is endowed with the canonical integration
based on the measure $\mu$, where
\begin{equation}\label{eq.7}
d\mu(q)=\sqrt{|det[G_{ab}]|}\;dq^{1}\ldots dq^{f},
\end{equation}
$f$ denoting the number of degrees of freedom ($n(n+1)/2$
in our model). Wave functions are elements of Hilbert space
${\rm L}^{2}(Q,\mu)$, $Q$ denoting the configuration space (here
$F(M,q)$), with the following scalar product:
\begin{equation}\label{eq.8}
\langle\psi_{1}|\psi_{2}\rangle=\int\overline{\psi_{1}}(q)\psi_{2}(q)d\mu(q).
\end{equation}

The kinetic energy is on the quantum level given by the operator
\begin{equation}\label{eq.9}
\widehat{T}=-\frac{\hbar^{2}}{2}\triangle, 
\end{equation}
where $\triangle$ is the Laplace-Beltrami operator given by
\begin{equation}\label{eq.10}
\triangle\psi=\frac{1}{\sqrt{|G|}}\sum_{ij}\frac{\partial}{\partial q^{i}}\left(\sqrt{\left|G\right|}G^{ij}\frac{\partial\psi}{\partial q^{j}}\right);
\end{equation}\label{eq.11}
$|G|$ is an abbreviation for $\det[G_{ab}]$, and, obviously, $G^{ij}$ denote the components of the contravariant reciprocal metric,
\begin{equation}\label{eq.12}
G^{ik}G_{kj}=\delta^{i}{}_{j}.
\end{equation}

Using the symbol $\nabla_{i}$
of the covariant derivative in the sense of the Levi-Civita connection induced by $G$, we can write down $\Delta\Psi$ in the following concise way
\begin{equation}\label{eq.13}
\Delta\Psi=G^{ij}\nabla_{i}\nabla_{j}\psi.
\end{equation}

The Hamilton operator $\widehat{H}$ is given by 
\begin{equation}\label{eq.14}
\widehat{H}=\widehat{T}+V,
\end{equation}
where $V$ denotes the potential energy.

\section{Some examples in two-dimensional constant-curvature
spaces.}

We present the quantization of classical two-dimensional problems
considered in \cite{Gol01, Gol02, Gol03, Gol04S, Prace IPPT_1, Prace IPPT_2, ROMP_SG}, i.e., the infinitesimal two-dimensional
gyroscope in the sphere and Lobatschevski pseudosphere.

In the spherical case the manifold is parameterized by polar variables 
$(r,\phi)$, $r\in [0,\pi R]$, $\phi\in [0,2\pi]$ and the metric element is
given by
\begin{equation}\label{eq.15}
ds^{2}=dr^{2}+R^{2}\sin^{2}\left(\frac{r}{R}\right)\,d\phi^{2}.
\end{equation}

Obviously, this is the restriction of the Euclidean metric element in $\mathbb{R}^{3}$,
\begin{equation}\label{eq.16}
dS^{2}=dx^{2}+dy^{2}+dz^{2},
\end{equation}
to the sphere of radius $R$, $S^{2}(0,R)\subset \mathbb{R}^{3}$, given parametrically by equations
\begin{eqnarray}\label{eq.17}
x&=&R\,\sin\left(\frac{r}{R}\right)\cos\phi, \nonumber \\
y&=&R\,\sin\left(\frac{r}{R}\right)\sin\phi,  \\
z&=&R\,\cos\left(\frac{r}{R}\right). \nonumber 
\end{eqnarray}

Obviously, $r$ is the distance from the ``north pole'' measured along the ``meridian'' (because of this the ``latitude'' $\theta=r/R$ runs over range $[0,\pi]$; $\theta=\pi$ corresponds to the ``south pole''), $\phi$ measured along ``equator'' is the ``longitude''. The Riemannian scalar curvature is given by $2/R^2$. The factor 2 or its absence is a matter of convention in the definition of Riemann tensor.

The basic orthonormal frame is given by:
\begin{eqnarray}
E_{r}&=&\frac{\partial}{\partial r}=\epsilon_{r},\label{eq.18}\\
E_{\phi}&=&\frac{1}{R\sin \left(\frac{r}{R}\right)}\frac{\partial}{\partial
\phi}=\frac{1}{R\sin \left(\frac{r}{R}\right)}\,\epsilon_{\phi}.\nonumber
\end{eqnarray}
Analytically, in terms of components:
\begin{eqnarray}
E_{r}&=&\left[\begin{array}{c}
  1  \\
  0 
\end{array} \right],\label{eq.19}\\
E_{\phi}&=& \left[\begin{array}{c}
 0   \\
  \frac{1}{R\sin\left(\frac{r}{R}\right)}
\end{array} \right].\nonumber                   
\end{eqnarray}
The script symbols $\epsilon_{r}$, $\epsilon_{\phi}$ denote the basic vectors tangent to coordinate lines. These vectors are orthogonal, but non-normalized to unity. The above normalization leads to the frame $\left(E_{r},E_{\phi}\right)$. Attention: the latter frame is non-holonomic, in spite of being collinear with holonomic frame $\epsilon_{r}$, $\epsilon_{\phi}$.

Rotations of the moving frame $(e_{1},e_{2})$ are parameterized by the
variable $\Psi$ where the orthogonal matrix $\left[\begin{array}{cc}
\cos\Psi & \sin\Psi\\
-\sin\Psi & \cos\Psi\\
\end{array}
\right]$ 
produces the current orientation of the body-fixed axes from the reference
frame as follows:
\begin{eqnarray}
e_{1}&=&E_{r}\cos\psi+E_{\phi}\sin\psi,\label{eq.20}\\
e_{2}&=&-E_{r}\sin\psi+E_{\phi}\cos\psi.\nonumber
\end{eqnarray}

The classical kinetic energy is given by \cite{Burov-Stepanov, VK_SSJ, Sl_1982, Sl-Sl, JJS_VK1, Sl_2002, SSJ_VK, Raman, Schredinger, SO4, ROZ_2010, ROZ_2004}:
\begin{equation}\label{eq.21}
T=T_{\rm tr}+T_{\rm rot}=\frac{M}{2}\left(\left(\frac{dr}{dt}\right)
^{2}+R^{2}\sin^{2}\left(\frac{r}{R}\right)\left(\frac{d\phi}{dt}\right)^{2}
\right)+\frac{I}{2}\,\Omega^{2},
\end{equation}
where $M$ denotes mass and $I$ is the scalar moment of inertia. The
labels ``tr'', ``rot'' refer respectively to translational and
rotational, i.e., internal motion. The symbol $\Omega$ denotes the
only independent component of the matrix of angular velocity,
\begin{equation}\label{eq.22}
\left[\Omega^{AB}\right]=\left[\begin{array}{cc}
 0 &  -\Omega    \\
 \Omega &  0  
\end{array} \right],
\end{equation}
this is the peculiarity of the spatial dimension $n=2$. It may be easily shown that
\begin{equation}\label{eq.23}
\Omega=\frac{d\Psi}{dt}+\cos\left(\frac{r}{R}\right)\frac{d\phi}{dt}.
\end{equation}
It is important that in addition to the naively expected term $d\Psi/dt$, there is an additional one, $\cos(r/R)(d\phi/dt)$ which, roughly speaking, describes the rotation of the frame $E$ itself.

It may be convenient to represent $T$ as follows:
\begin{eqnarray}
T&=&\frac{M}{2}\,G_{ij}\frac{dq^{i}}{dt}
\frac{dqj}{dt}=\label{eq.24}\\
&=&\frac{m}{2}\left(\left(\frac{dr}{dt}\right)^{2}+
R^{2}\sin^{2}\left(\frac{r}{R}\right)\left(\frac{d\phi}{dt}\right)^{2}\right)+
\frac{I}{2}\left(\frac{d\Psi}{dt}+\cos\left(\frac{r}{R}\right)\frac{d\phi}{dt}\right)^{2},\nonumber
\end{eqnarray}
where the particular shape of $T_{\rm tr}$, $T_{\rm rot}$ is specified by
\begin{eqnarray}
T&=&T_{\rm tr}+T_{\rm rot}=\label{eq.25}\\
&=&
\frac{M}{2}\left(\left(\frac{dr}{dt}\right)^{2}+
R^{2}\sin^{2}\left(\frac{r}{R}\right)\left(\frac{d\phi}{dt}\right)^{2}\right)+
\frac{I}{2}\left(\frac{d\theta}{dt}+\cos\left(\frac{r}{R}\right)\frac{d\phi}{dt}\right)^{2}.\nonumber
\end{eqnarray}
In the pseudospherical case we use coordinates $(r,\phi)$, where
now $r\in [0,\infty]$,
\begin{equation}\label{eq.26}
ds^{2}=dr^{2}+R^{2}\,{\rm sh}^{2}\left(\frac{r}{R}\right)\:d\phi^{2}.
\end{equation}
Again, this is the restriction of the pseudo-Euclidean metric element in $\mathbb{R}^{3}$,
\begin{equation}\label{eq.27}
dS^{2}=dx^{2}+dy^{2}-dz^{2},
\end{equation}
to the pseudosphere, i.e., hyperboloid $H^{2}(0,R)\subset\mathbb{R}^{3}$ of the pseudoradius $R$,
\begin{eqnarray}
x&=&R\ {\rm sh}\left(\frac{r}{R}\right)\:\cos\phi, \nonumber \\
y&=&R\ {\rm sh}\left(\frac{r}{R}\right)\:\sin\phi, \label{eq.28}\\
z&=&R\ {\rm ch}\left(\frac{r}{R}\right). \nonumber
\end{eqnarray}
Again, $r$ is the distance from the ``pole'' $(0,0,R)$ measured along the 
geodesic (hyperbole).

\noindent {\bf Remark:} in both cases, i.e., spherical and pseudospherical (hyperbolic, Lobatschevski), the reference frames $(E_{r},E_{\phi})$ are non-holonomic, because the components of $\partial/\partial\phi$ depend on $r$. One point is important, namely these nonholonomic frames are singular at the ``pole'' $r=0$. In the spherical case the singularity in unavoidable, because $S^{2}(0,R)$ is non-parallelizable. However if properly and carefully treated, this singularity is harmless, just as one of polar coordinates in $\mathbb{R}^{2}$. The basic orthonormal frame is given by
\begin{eqnarray}
E_{r}&=&\frac{\partial}{\partial r}=\epsilon_{r},\label{eq.29}\\
E_{\phi}&=&\frac{1}{R\,{\rm sh}\left(\frac{r}{R}\right)}\frac{\partial}{\partial \phi}=\frac{1}{R\, {\rm sh}\left(\frac{r}{R}\right)}\,\epsilon_{\phi},\nonumber
\end{eqnarray}
similarly like in (\ref{eq.18}),(\ref{eq.19}). And again, in terms of analytical, component-wise expressions we have that
\begin{eqnarray}
E_{r}&=&\left[\begin{array}{c}
  1   \\
  0
\end{array} \right],\label{eq.30}\\ 
E_{\phi}&=&\left[\begin{array}{c}
  0   \\
  \frac{1}{R\,{\rm sh}\left(\frac{r}{R}\right)}
\end{array} \right]\nonumber
\end{eqnarray}
and the frame $(E_{r},E_{\phi})$ is non-holonomic, unlike $(\epsilon_{r},\epsilon_{\phi})$.

The kinetic energy is given by
\begin{eqnarray}
T&=&T_{\rm tr}+T_{\rm rot}=\label{eq.31}\\
&=&\frac{M}{2}\left(\left(\frac{dr}{dt}\right)^{2}+
R^{2}\:{\rm sh}^{2}\left(\frac{r}{R}\right)\left(\frac{d\phi}{dt}\right)^{2}\right)+
\frac{I}{2}\left(\frac{d\Psi}{dt}+{\rm ch}\left(\frac{r}{R}\right)\frac{d\phi}{dt}\right)^{2}.\nonumber
\end{eqnarray}
This time we have that
\begin{equation}\label{eq.32}
\Omega=\frac{d\Psi}{dt}+{\rm ch}\left(\frac{r}{R}\right)\frac{d\phi}{dt}.
\end{equation}
In the common form:
\begin{equation}\label{eq.33}
T=\frac{M}{2}\,G_{ij}\frac{dq^{i}}{dt}\frac{dq^{i}}{dt},
\end{equation}
where $\left(q^{1}, q^{2}, q^{3}\right)=\left(r,\phi,\Psi\right)$ are generalized coordinates. The quantized kinetic energy is built of the Laplace-Beltrami operator \cite{MaxBorn}:
\begin{equation}\label{eq.34}
\widehat{T}=-\frac{\hbar^{2}}{2M}\Delta=-\frac{\hbar^{2}}{2M}\frac{1}{\sqrt{\left|G\right|}}\sum_{ij}\frac{\partial}{\partial q^i}\sqrt{\left|G\right|}\,G^{ij}\frac{\partial}{\partial q^j}.
\end{equation}

The spherical problem is isomorphic with the 3-dimensional symmetric rigid body without translational motion, i.e., with some left-invariant metric on ${\rm SO}(3, \mathbb{R})$ invariant also on right under rotations around the z-axis. Similarly, the pseudospherical problem is isomorphic with the corresponding problem on the Lorentz group ${\rm SO}(1,2)$, locally isomorphic with ${\rm SL}(2, \mathbb{R})$.

The Laplace-Beltrami operator in the spherical case has the following form:
\begin{eqnarray}
\Delta&=&\frac{\partial^{2}}{\partial r^{2}}+\frac{1}{R}\,{\rm ctg}\left(\frac{r}{R}\right)\frac{\partial}{\partial r}+\frac{1}{R^{2}\sin^{2}\left(\frac{r}{R}\right)}\frac{\partial^{2}}{\partial\phi^{2}}+\label{eq.35}\\
&-&\frac{2\cos\left(\frac{r}{R}\right)}{R^{2}\sin^{2}\left(\frac{r}{R}\right)}
\frac{\partial^{2}}{\partial\phi\partial\Psi}+\frac{MR^{2}\sin^{2}(r/R)+
I\cos^{2}\left(\frac{r}{R}\right)}{IR^{2}\sin^{2}\left(\frac{r}{R}\right)}\frac{\partial^{2}}
{\partial\Psi^{2}}.\nonumber
\end{eqnarray}
In particular, in the very special case, when $I=MR^{2}$, we obtain that
\begin{eqnarray}
\Delta_{0}&=&\frac{\partial^{2}}{\partial r^{2}}+\frac{1}{R}\,{\rm ctg}\left(\frac{r}{R}\right)\frac{\partial}{\partial r}+\frac{1}{R^{2}\sin^{2}\left(\frac{r}{R}\right)}\frac{\partial^{2}}{\partial\phi^{2}}+\label{eq.36}
\\
&-&\frac{2\cos\left(\frac{r}{R}\right)}{R^{2}\sin^{2}\left(\frac{r}{R}\right)}\frac{\partial^{2}}
{\partial\phi\partial\Psi}+\frac{1}{R^{2}\sin^{2}\left(\frac{r}{R}\right)}
\frac{\partial^{2}}{\partial\Psi^{2}}.\nonumber
\end{eqnarray}
This problem is isomorphic with the spherical 3-dimensional top
without translational motion, i.e., with some left-invariant metric
on ${\rm SO}(3,\mathbb{R})$ invariant also on right under rotations about
the z-axis.

In the pseudospherical problem trigonometric functions are replaced by
hyperbolic ones, and indeed, after some calculations one obtains the
following expression
\begin{eqnarray}
\Delta&=&\frac{\partial^2}{\partial r^2}+\frac{1}{R}\,{\rm cth}\left(\frac{r}{R}\right)\frac{\partial}{\partial r}-\frac{2\,{\rm ch}\left(\frac{r}{R}\right)}{R^2\,{\rm sh}^2\left(\frac{r}{R}\right)}\frac{\partial^2}
{\partial\psi\partial\phi}+\label{eq.37}\\
&+&\frac{1}{R^2\,{\rm sh}^2\left(\frac{r}{R}\right)}\frac{\partial^2}{\partial\phi^{2}}+
\frac{MR^2\,{\rm sh}^2\left(\frac{r}{R}\right)+I\,{\rm ch}^2\left(\frac{r}{R}\right)}{IR^2\,{\rm sh}^2\left(\frac{r}{R}\right)}\frac{\partial^2}{\partial\psi^2}.\nonumber
\end{eqnarray}

It is seen that really, as expected, the classical and quantum formulas
for the spherical and pseudospherical geometry may be quite formally
obtained from each other by the simple interchanging between trigonometric
and hyperbolic functions. Of course, the geometry and topology of the
corresponding configuration spaces are quite different. Formally this is
also reflected by the different ranges of the r-variable: $[0,\pi R]$
in the spherical case, and $[0,\infty]$ in hyperbolic geometry.

There is one interesting point concerning similarities and
differences between both models. Formally, the configuration space of
infinitesimal rigid body in the spherical space, may be identified
with the three-dimensional rotation group ${\rm SO}(3,\mathbb{R})$.
Configuration space of translational degrees of freedom, $S^{2}(0,R)$
may be identified with the quotient manifold
${\rm SO}(3,\mathbb{R})/{\rm SO}(2,\mathbb{R})$. The ``denominator"
${\rm SO}(2,\mathbb{R})$ stands here for the group of planar rotations
about the "z-axis" in $\mathbb{R}^{3})$. And this one-dimensional
group represents here internal, gyroscopic degrees of freedom. The
problem is formally isomorphic with the mechanics of
three-dimensional symmetric top in $\mathbb{R}^{3}$ (without
translational degrees of freedom). The two main moments of its
three-dimensional inertia are $MR^{2}$ and $I$, $MR^{2}$ being degenerate eigenvalue of the inertial matrix. This is seen when $r$ is expressed by the
angular variable $\theta$,
\begin{equation}\label{eq.38}
r=R\theta.
\end{equation}

From the three-dimensional point of view $(\phi,\theta,\psi)$ are then Euler angles, namely $\phi$ is the precession angle, $\theta$ the nutation angle, and $\psi$ is the proper rotation. From the mentioned ``geographical'' perspective of the two-dimensional sphere, $(\phi,\theta,\psi)$ are in a sense ``longitude'', ``latitude'', ``altitude'' respectively. These angles correspond to the representation of orthogonal matrices $U\in {\rm SO}(3,\mathbb{R})$ as products:
\begin{eqnarray}
&&U(\phi,\theta,\psi)=\label{eq.39}\\
&&=\left[\begin{array}{ccc}
 \cos\phi & -\sin\phi  & 0  \\
 \sin\phi & \cos\phi  &  0 \\
 0 &  0 & 1
\end{array} \right]
\left[\begin{array}{ccc}
 1 & 0  & 0  \\
 0 & \cos\theta  & -\sin\theta  \\
 0 & \sin\theta  & \cos\theta
\end{array} \right]
\left[\begin{array}{ccc}
 \cos\psi & -\sin\psi  & 0  \\
 \sin\psi & \cos\psi  & 0  \\
 0 & 0  & 1
\end{array} \right].\nonumber
\end{eqnarray}
The matrix of co-moving angular velocity has the following form: 
\begin{equation}\label{eq.40}
\widehat{\omega}=U^{-1}\frac{dU}{dt}=U^T\frac{dU}{dt}= \left[\begin{array}{ccc}
0& -\omega_{3} & \omega_{2} \\
\omega_{3} & 0 & -\omega_{1} \\
-\omega_{2} & \omega_{1} & 0
\end{array} \right],
\end{equation}
where
\begin{eqnarray}
\omega_{1}&=&\cos\psi \, \frac{d\theta}{dt}+\sin\theta\sin\psi \, \frac{d\phi}{dt},
\nonumber\\
\omega_{2}&=&-\sin\psi \, \frac{d\theta}{dt}+\sin\theta\cos\psi \, \frac{d\phi}{dt},\label{eq.41}
\\
\omega_{3}&=&\frac{d\psi}{dt}+\cos\theta \, \frac{d\phi}{dt}.\nonumber
\end{eqnarray}
The corresponding ``kinetic energy'' has the following form:
\begin{equation}\label{eq.42}
T=\frac{I_{1}}{2}(\omega_{1})^2+\frac{I_{2}}{2}(\omega_{2})^2+
\frac{I_{3}}{2}(\omega_{3})^2,
\end{equation}
where
\begin{equation}\label{eq.43}
I_{1}=I_{2}=M{}R^2, \qquad I_{3}=I.
\end{equation}
This is exactly (\ref{eq.42}) with $r=R\theta$.

The kinetic energy (\ref{eq.25}) and the corresponding metric tensor on ${\rm SO}(3,\mathbb{R})$  are invariant under all transformations: 
\begin{equation}\label{eq.44}
{\rm SO}(3,\mathbb{R})\ni U\mapsto AU\in {\rm SO}(3,\mathbb{R}),\qquad 
A\in {\rm SO}(3,\mathbb{R}), 
\end{equation}
i.e., under all left regular translations in ${\rm SO}(3,\mathbb{R})$. It is also invariant under right regular translations by rotations about the ``z-axis'', i.e., by the corresponding right action of 
\begin{equation}\label{eq.45}
{\rm SO}(3,\mathbb{R})\ni U\mapsto UA\in {\rm SO}(3,\mathbb{R}),\quad
A=\left[\begin{array}{cc}
B & 0 \\
0 & 1 
\end{array} \right],\quad B^T B=I_{2},
\end{equation}
and $B\in {\rm SO}(2,\mathbb{R})$ are $2\times2$ orthogonal matrices.   

It is easy to see that (\ref{eq.25}),(\ref{eq.42}) is perfectly invariant under ${\rm SO}(3,\mathbb{R})\times {\rm SO}(3,\mathbb{R})$, i.e., under all left  and right translations in ${\rm SO}(3,\mathbb{R})$, when $I=M{}R^2$, i.e., when the corresponding ``three-dimensional top'' is spherical. The natural question arises as to the analogous  problems in pseudosphere (hyperbolic) geometry. Something similar may be formulated then, nevertheless, certain essential differences appear due to the non-compactness of the pseudosphere. Namely, it is seen that the special case $I=M{}R^2$ in (\ref{eq.31}) is not so particular and simplifying like in (\ref{eq.25}), (\ref{eq.42}). The reason is that it does not correspond to the maximal possible symmetry. Let  us explain this briefly. The hyperbole (\ref{eq.28}) may be parameterized in a form similar to Euler angles, however with one ``hyperbolic'' coordinate. The hyperbole $H^2(0,R)$ may be identified with the quotient manifold ${\rm SO}(1,2)/{\rm SO}(2,\mathbb{R})$  of the three-dimensional Lorentz group ${\rm SO}(1,2)$ with respect to the group of planar rotations. We mean the Lorentz group ${\rm SO}(1,2)$ of $\mathbb{R}^3$ preserving the metric (\ref{eq.27}); ${\rm SO}(2,\mathbb{R})$ is the group of rotations in the $(x,y)$-plane, preserving the variable $z$. So, for any $L\in {\rm SO}(1,2)$ we have the representation analogous to Euler angles:
\begin{eqnarray}
&&L(\phi,\chi,\psi)=\label{eq.46}\\
&&\left[\begin{array}{ccc}
\cos\phi &-\sin\phi & 0 \\
\sin\phi & \cos\phi & 0 \\
0 & 0 & 0
\end{array} \right]
\left[\begin{array}{ccc}
 1 & 0  & 0  \\
 0 & {\rm ch}\chi  & {\rm sh}\chi  \\
 0 & {\rm sh}\chi  & {\rm ch}\chi
\end{array} \right]
\left[\begin{array}{ccc}
 \cos\psi & -\sin\psi  & 0  \\
 \sin\psi & \cos\psi  &  0 \\
 0 & 0  & 1
\end{array} \right],\nonumber
\end{eqnarray}
where $\chi=r/R$, the meaning of $r$ as previously. Then for the ``pseudo-angular  velocity''
\begin{equation}\label{eq.47}
\lambda=L^{-1}\frac{dL}{dt}
\end{equation}
we have that
\begin{equation}\label{eq.48}
\lambda=\left[\begin{array}{ccc}
 0 & -\lambda_{3}  & \lambda_{2}  \\
 \lambda_{3} & 0  & \lambda_{1}  \\
 \lambda_{2} & \lambda_{1}  & 0
\end{array} \right],
\end{equation}
where
\begin{eqnarray}
\lambda_{1}&=&\cos\psi \,\frac{d\chi}{dt}+{\rm sh}\chi \sin\psi \,\frac{d\phi}{dt},\nonumber\\
\lambda_{2}&=&\sin\psi \,\frac{d\chi}{dt}-{\rm sh}\chi \cos\psi \,\frac{d\phi}{dt},\label{eq.49}\\
\lambda_{3}&=&\frac{d\psi}{dt}+{\rm ch}\chi \,\frac{d\phi}{dt}.\nonumber
\end{eqnarray}
In analogy to (\ref{eq.42}), the kinetic energy (\ref{eq.31}) may be written down as follows:
\begin{equation}\label{eq.50}
T=\frac{I_{1}}{2}(\lambda_{1})^2+\frac{I_{2}}{2}(\lambda_{2})^2+\frac{I_{3}}{2}(\lambda_{3})^2,
\end{equation}
where
\begin{equation}\label{eq.51}
I_{1}=I_{2}=MR^2, \qquad I_{3}=I.
\end{equation}

As mentioned, the special case $I=M{}R^2$, although simplifies some expressions is not geometrically so distinguished and peculiar as in the case of spherical geometry. In any case (\ref{eq.50}) and its quantum counterpart (\ref{eq.37}) do not become then invariant under the total ${\rm SO}(1,2)$ acting on the right on itself, and therefore, on the configuration space of our object (rotator moving in a hyperbolic space). The point is that in the case of the hyperbolic signature $(+ + -)$ the corresponding expressions with $I=MR^2$ are not Casimir invariants. The Casimir invariant on ${\rm SO}(1,2)$ corresponds to taking $I_{1}=I_{2}=-I_{3}=I$. Obviously, one is rather reluctant to ``kinetic energy'' which fails to be positively definite. However, although rather unacceptable in the very fundamental physical theories (matter would be unstable), they may be useful in certain viable dynamical models in applications. In any case, they are interesting at least from the geometric point of view. Because of this it may be convenient to discuss also some academic models with the negative contribution of rotations in the kinetic energy term. Then (\ref{eq.31}) is replaced by
\begin{equation}\label{eq.52}
T^{(-)}=\frac{M}{2}\left(\left(\frac{dr}{dt}\right)^2+
R^2{\rm sh}^2\left(\frac{r}{R}\right)\left(\frac{d\phi}{dt}\right)^{2}\right)-
\frac{I}{2}\left(\frac{d\psi}{dt}+{\rm ch}\left(\frac{r}{R}\right)\frac{d\psi}{dt}\right)^2
\end{equation}
and (\ref{eq.37}) by
\begin{equation}\label{eq.53}
\widehat{T}^{(-)}=-\frac{\hbar^2}{2M}\Delta^{(-)},
\end{equation}
where
\begin{eqnarray}
\Delta^{(-)}&=&\frac{\partial^2}{\partial r^2}+\frac{1}{R}\,{\rm cth}\left(\frac{r}{R}\right)\frac{\partial}{\partial r}-\frac{2{\rm ch}\left(\frac{r}{R}\right)}{R^2\,{\rm sh}^2\left(\frac{r}{R}\right)}\frac{\partial^2}{\partial\psi\partial\phi}\label{eq.54}\\
&+&\frac{1}{R^2\,{\rm sh}^2\left(\frac{r}{R}\right)}\frac{\partial^2}{\partial\phi^2}+
\frac{I{\rm ch}^2\left(\frac{r}{R}\right)-MR^2\,{\rm sh}^2\left(\frac{r}{R}\right)}{IR^2\,{\rm sh}^2\left(\frac{r}{R}\right)}\frac{\partial^2}{\partial\psi^2}.\nonumber
\end{eqnarray}

The radical geometric simplification of $T^{(-)}$, $\Delta^{(-)}$ and their high invariance under ${\rm SO}(1,2)\times {\rm SO}(1,2)$ when $I=M{}R^2$ are obvious. We consider parallelly both models. It is convenient to denote (\ref{eq.31}) by $T^{(+)}$ and (\ref{eq.37}) by $\widehat{T}^{(+)}=-\left(\hbar^2/2M\right)\Delta^{(+)}$ to distinguish explicitly between the elliptic signature in (\ref{eq.31}) and the normal-hyperbolic signatures (\ref{eq.52}). What concerns the invariant Riemannian measure, in the spherical and pseudospherical case we have respectively:
\begin{eqnarray}
\sqrt{\left|G\right|}&=&\sqrt{\frac{I}{M}}\,R\sin\left(\frac{r}{R}\right) \qquad (\rm spherical),\label{eq.55}\\
\sqrt{\left|G\right|}&=&\sqrt{\frac{I}{M}}\,R\,{\rm sh}\left(\frac{r}{R}\right) \qquad (\rm pseudospherical).\label{eq.56}
\end{eqnarray}

The expression (\ref{eq.36}) remains valid when the positive signature $(+++)$ of the configuration space is replaced by the normal-hyperbolic one $(++-)$ (the negative configuration of internal rotations to the ``kinetic energy'' expression). Obviously, the constant factors in (\ref{eq.55}), (\ref{eq.56}) are non-essential. In the compact spherical case, two normalizations are often used: such that the total volume equals one (convention typical for group theory), or that it equals just the volume (three-dimensional ``area'') of the sphere of radius $R$, $S^3(0,R)$ in the four-dimensional Euclidean space $\mathbb{R}^4$.

\section{Injected torus in $\mathbb{R}^3$ as the configuration space}

The above examples, i.e., quantum mechanics of the rotator in the spherical and pseudospherical spaces $S^2(0,R)$, $H^2(0,R)$ are geometrically special, because these manifolds are constant-curvature spaces. They have three-dimensional groups of motion-the maximal dimension of the isometry group in two dimensions. Both the classical and quantum problems may be interpreted in terms of invariant dynamical systems on Lie groups ${\rm SO}(3,\mathbb{R})$, ${\rm SO}(1,2)$, i.e., orthogonal group and Lorentz group in $\mathbb{R}^3$. They may be also considered in groups ${\rm SU}(2)$ (the universal covering of ${\rm SO}(3,\mathbb{R})$) and ${\rm SL}(2,\mathbb{R})$ which is also locally isomorphic with ${\rm SO}(1,2)$.

Both manifolds, i.e., sphere and pseudosphere are algebraic submanifolds of $\mathbb{R}^{3}$. As algebraic manifolds they have degree 2, because they are given as value-surfaces of second-order polynomials respectively: 
\begin{equation}\label{eq.57}
x^2+y^2+z^2=R^{2}, \qquad -x^2-y^2+z^2=R^{2}.
\end{equation}

In the hyperbolic case we must take the fixed sign of $z$, e.g., $z>{0}$ (as a matter of fact, $z\geq R$) if the manifold is to be connected (one shell of the two-shell hyperboloid).

There is also another interesting example where some calculations may be effectively carried out, namely the injected two-dimensional torus in $\mathbb{R}^{3}$, with the metric induced by the restriction of the surrounding 
euclidean metric. Geometric properties are weaker now, namely, there is only one-dimensional group of isometries, and the curvature is not constant. In this sense, the flat Euclidean torus obtained as a quotient of $\mathbb{R}^3$ modulo $\mathbb{Z}^3$ is geometrically perhaps more interesting.  Nevertheless, motion along the injected torus may be interesting from the point of view of applications. And mathematically, the injected torus is also an algebraic manifold, this time one of degree four. It is natural to expect that classical and quantum systems on algebraic manifolds are somehow interesting, perhaps also from the point of view of integrability.  

The injected two-dimensional torus in three-dimensional Euclidean space,
$T^2(0;L,R)$ $\subset \mathbb{R}^3$ is a parametrically given by
\begin{eqnarray}
x&=&\left(L+R\cos\theta\right)\cos\phi,  \nonumber\\
y&=&\left(L+R\cos\theta\right)\sin\phi,  \label{eq.58}\\
z&=&R\sin\theta,\nonumber
\end{eqnarray}
where $L$, $R$ are constants, and $\phi,\theta$ are respectively geographical ``longitude" and ``latitude". What concerns ``latitude", the range is different than that in spherical geometry. Namely, the both angular variables $\phi$, $\theta$ parameterizing $T^2(0;L,R)$ run over the range $[0,2\pi]$ with the obvious "identification" of $2\pi$ with $0$. $L$ is the ``large" radius, i.e., the radius of the centrally placed inner-"tube" circle. $R$ is the ``small" radius, i.e., the radius of circles which are obtained as intersections of the ``tube" by planes passing through the ``$z$-axis". The origin of coordinates in $\mathbb{R}^3$, $(0,0,0)$ is placed centrally with respect to the whole figure. As mentioned, $T^2(0;L,R)$ is a fourth-degree algebraic  surface in $\mathbb{R}^3$. The corresponding equation has the following form:
\begin{equation}\label{eq.59}
\left(x^2+y^2+z^2+L^2-R^2\right)^2-4L^2\left(x^2+y^2\right)=0.
\end{equation}
After substituting to the 3-dimensional Euclidean metric $dx^2+dy^2+dz^2$, we obtain the first quadratic form of $T^2(0;L,R)$, i.e., the metric element
\begin{equation}\label{eq.60}
ds^2=R^2d\theta^2+\left(L+R\cos\theta\right)^2d\phi^2.
\end{equation}

In analogy to the spherical and pseudospherical geometry, it is sometimes convenient to use the variable
\begin{equation}\label{eq.61}
r=R{}\theta{}\in[0,2\pi R],
\end{equation}
i.e., the ``distance'' measured along ``meridians''. Then we have that
\begin{equation}\label{eq.62}
ds^2=dr^2+\left(L+R\cos\left(\frac{r}{R}\right)\right)^2d\phi^2.
\end{equation}

Kinetic energy of the material point moving on $T^2(0;L,R)$ has the form:
\begin{eqnarray}
T_{\rm tr} & = & \frac{M}{2}\left(\left(\frac{dr}{dr}\right)^2+
\left(L+R\cos\left(\frac{r}{R}\right)\right)^2\left(\frac{d\phi}{dt}\right)^2\right)=
\nonumber\\
&=& \frac{M}{2}\left(R^2\left(\frac{d\theta}{dt}\right)^2+
\left(L+R\cos\left(\frac{r}{R}\right)\right)^2\left(\frac{d\phi}{dt}\right)^2\right)=\label{eq.63}\\
&=&
\frac{M{}R^2}{2}\left(\left(\frac{d\theta}{dt}\right)^2+\left(\frac{L}{R}+
\cos\theta \right)^2\left(\frac{d\phi}{dt}\right)^2\right).\nonumber
\end{eqnarray}
The corresponding two-dimensional volume, i.e., area element on $T^2(0;L,R)$ is given by
\begin{equation}\label{eq.64}
d\mu=R\left(L+R\cos\theta\right)d\theta d\phi=\left(L+R\cos\left(\frac{r}{R}\right)\right)drd\phi.
\end{equation}
Metric tensor $g$ and its contravariant inverse $g^{-1}$ have respectively the matrices:
\begin{equation}\label{eq.65}
\left[g_{ij}\right]=\left[
\begin{array}{cc}
 R^2 & 0     \\
 0 & (L+R\cos\theta)^2    
\end{array} 
\right],\qquad 
\left[g^{ij}\right]=\left[
\begin{array}{cc}
 \frac{1}{R^2} & 0     \\
 0 & \frac{1}{(L+R\cos\theta)^2}     
\end{array}
\right],
\end{equation}
and
\begin{equation}\nonumber
\sqrt{\left|g\right|}=R(L+R\cos\theta).
\end{equation}

The auxiliary field of orthonormal frames will be chosen as follows:
\begin{eqnarray}
E_{\theta}&=&\frac{1}{R}\frac{\partial}{\partial\theta}=
\frac{1}{R}\,\epsilon_{\theta},\label{eq.66}\\
E_{\phi}&=&\frac{1}{L+R\cos\theta}\frac{\partial}{\partial\phi}=
\frac{1}{L+R\cos\theta}\,\epsilon_{\phi}.\nonumber
\end{eqnarray}
Now, let the material point with the mass $M$ be replaced by the infinitesimal rotator with the mass $M$ and the inertial moment $I$.

The kinetic energy may be then expressed as
\begin{eqnarray}
T&=&T_{\rm tr}+T_{\rm int}=\label{eq.67}\\
&=&\frac{M}{2}\left(R^2\left(\frac{d\theta}{dt}\right)^2+
\left(L+R\cos\theta\right)^2\left(\frac{d\phi}{dt}\right)^2\right)+
\frac{I}{2}\left(\frac{d\psi}{dt}+\sin\theta\frac{d\phi}{dt}\right)^2.\nonumber
\end{eqnarray}
It may be written in the following form:
\begin{equation}\label{eq.68}
T=\frac{M}{2}\,G_{ij}(q)\frac{dq^i}{dt}\frac{dq^j}{dt},
\end{equation}
where generalized coordinates are ordered as
\begin{equation}\label{eq.69}
\left(q^1,q^2,q^3\right)=\left(\theta,\phi,\psi\right).
\end{equation}

Metric tensor $G$ underlying the kinetic energy (\ref{eq.67}), (\ref{eq.68}) has the matrix 
\begin{equation}\label{eq.70}
[G_{ij}]=\left[\begin{array}{ccc}
 R^2 & 0 & 0  \\
 0 & \left(L+R\cos\theta\right)^2+\frac{I}{M}\sin^2\theta  & \frac{I}{M}\sin\theta  \\
 0 & \frac{I}{M}\sin\theta  & \frac{I}{M}
\end{array} \right].
\end{equation}
The contravariant inverse has the following matrix:
\begin{equation}\label{eq.71}
[G^{ij}]=\left[\begin{array}{ccc}
 \frac{1}{R^2} & 0 & 0  \\
 0 & \frac{1}{(L+R\cos\theta)^2} & -\frac{\sin\theta}{(L+R\cos\theta)^2}  \\
 0 & -\frac{\sin\theta}{(L+R\cos\theta)^2}  & \frac{M}{I}+\frac{\sin^2\theta}{(L+R\cos\theta)^2}.
\end{array} \right]
\end{equation}
The corresponding volume element in the configuration space has the form:
\begin{equation}\label{eq.72}
d{\rm Vol}(\phi,\theta,\psi)=\sqrt{\left|G \right|}\,d\phi d\theta d\psi=\sqrt{\frac{I}{M}}\,R(L+R\cos\theta).
\end{equation}

Let us turn to the quantum problem. The classical Hamiltonian
\begin{equation}\label{eq.73}
H=T+V=\frac{1}{2}\,G^{ij}p_i p_j +V
\end{equation}
is replaced by the operator
\begin{equation}\label{eq.74}
\widehat{H}=-\frac{\hbar^2}{2M}\Delta_G +V,
\end{equation}
where $\Delta_G$ is the Laplace-Beltrami operator built of the metric tensor.

\section{Separation of variables}

We have discussed above the general structure of three models with two-dimen\-sional algebraic manifolds used as the configuration space of translational motion. The total configuration spaces of the "small", infinitesimal, rigid body moving there, are fibre bundles over those two-dimensional spaces. They are principal fibre bundles of orthonormal frames. As the base manifolds are two-dimensional, the fibres are compact one-dimensional manifolds, i.e., circles. The structure groups is isomorphic with any of the groups ${\rm SO}(2,\mathbb{R})$, ${\rm U}(1)$; all compact one-dimensional Lie groups are structurally identical. Motion of the ``small'' top in spherical and pseudospherical manifolds is isomorphic with the motion of point in the groups ${\rm SO}(3,\mathbb{R})$, ${\rm SO}(1,2)$ respectively. Motion in the toroidal world has no such a type of group-theoretical background. We mean of course the injected torus in $\mathbb{R}^3$ with the induced curved metric. One should carefully distinguish two things here: topologically, the configuration space of gyroscopic motion in the two-dimensional world is identical with some problem in $\mathbb{T}^3=\mathbb{R}^3/\mathbb{Z}^3$ --- three-dimensional torus obtained as a quotient of $\mathbb{R}^3$ with respect to the integer lattice. But the flat metric structure in $\mathbb{T}^3$ is different than one based on the injected torus in $\mathbb{R}^3$. And the ``algebraic'' structure of $\mathbb{T}^3=\mathbb{R}^3/\mathbb{Z}^3$ is not ever used in our mechanical problem. This problem is not metrically isomorphic with the invariant Hamiltonian problem on the group 
\[
\mathbb{T}^3=\mathbb{R}^3/\mathbb{Z}^3\cong({\rm SO}(2,\mathbb{R}))^3\cong({\rm U}(1))^3. 
\]
The isometry group of injected torus is only one-dimensional.

In spite of the obvious differences, there exist, however, some similarities between three models. The variable $r=R{}\theta$ is structurally similar in all problems, in spite of topological differences. It is a distance from the ``north pole". Of course, in the pseudospherical world there is only one pole. If, by convention, we call it a north pole, then the ``south pole'' is in infinity. On the injected torus there is one-dimensional line of ``north poles'' $(\theta= 0)$ and one-dimensional line of ``south poles'' $(\theta=\pi)$. In the spherical world there are two polar points, the northern one, $(\theta=0)$ and the southern one $(\theta=\pi)$. Nevertheless, some similarities exist. In the kinetic energy of the metrical point in any of those three two-dimensional ``worlds'', the ``geographic longitude'' $\phi$ is a cyclic variable. In the internal part of gyroscopic motion, the ``altitude variable'' $\psi$ is also a cyclic coordinate. And even if there are no forces, the ``geographic latitude'' is an evidently non-cyclic coordinate. In all models the distance from the ``north'' is given by $r=R\theta$ (but, of course, one must be careful with range of $\theta$ in injected torus; it is different than in spherical world, namely, $r=0$ and $r=2{}\pi{}R$ represent the same situation --- the ``north'' pole; $r=\pi{}R$ is the ``south'' one).

When we use the natural, geometric coordinates $\left(\theta,\phi,\psi\right)$ (or, equivalently, $\left(r,\phi,\psi\right)$), then the matrices (\ref{eq.70}), (\ref{eq.71}) are non-diagonal. There is a linear theorem that in 3 dimensions any Riemannian metric is diagonal in appropriate coordinates. And the classical  separability theorems, like the Staeckal theorem, are suited just to metrics put in orthogonal, i.e., diagonal, form. However, in general, in particular for (\ref{eq.70}), (\ref{eq.71}) the corresponding formulas will be rather complicated, therefore, non-effective. But fortunately, we are interested rather in some special problems, when the explicit separability appears. As mentioned, in all three geodetic problems $\phi$, $\psi$ are cyclic coordinates. And similarly, the most natural class of potentials is one adapted to this structure of geodetic problems. The corresponding potentials are independent on the angles $\phi$, $\psi$ and are purely ``radial'', i.e., depending only on the distance $r$ from the ``north'' pole, i.e., equivalently, on the variable $\theta=r/R$. 

In classical problems the corresponding canonical momenta $P_{\phi}$, $P_{\psi}$ conjugate to $\phi$, $\psi$ are constants of motion. Then in the stationary Hamilton-Jacobi equation with energy value $E$,
\begin{equation}\label{eq.75}
\frac{1}{2M}\,G^{ij}\frac{\partial S}{\partial q^i}\frac{\partial S}{\partial q^j}+\textit{V}(q)=E,
\end{equation}
we substitute
\begin{equation}\label{eq.76}
S\left(q^1,q^2,q^3\right)=S\left(\theta,\phi,\psi\right)=
S_\theta (\theta)+\mu \phi+\sigma \psi,
\end{equation}
where $\mu$, $\sigma$ are constants, just the values of constants of motion $P_{\phi}$, $P_{\psi}$, fixed for a given trajectory. This results in ordinary differential equations for $S_{\theta}$ in a form whose integration is reducible to quadratures.

In quantum models we assume then the following separated form of the wave function
\begin{equation}\label{eq.77}
\Psi\left(q^1,q^2,q^3\right)=\Psi(\theta,\phi,\psi)
=f(\theta)e^{im\phi}e^{is\psi},
\end{equation}
where $m$, $s$ are integers-quantum numbers. They are eigenvalues of differential operators
\begin{equation}\label{eq.78}
\widehat{P}_\phi=\frac{\hbar}{i}\frac{\partial}{\partial\phi},\qquad
\widehat{P}_\psi=\frac{\hbar}{i}\frac{\partial}{\partial\psi}
\end{equation}
which physically represent respectively the ``orbital'' angular momentum of motion in the two-dimensional ``world'' and the ``spin'' angular momentum of internal rotation of the top. Therefore, substituting (\ref{eq.77}) into (78), we have that
\begin{equation}\label{eq.79}
\widehat{P}_\theta \Psi=\hbar m \Psi,\qquad 
\widehat{P}_\psi \Psi=\hbar s \Psi.
\end{equation}
When $m$, $s$ run over the set of integers, then $\hbar m$, $\hbar s$ take over the quantised values of quantities $\mu$, $\sigma$ in (\ref{eq.76}). Substituting (\ref{eq.77}) into the stationary Schr\"odinger equation,
\begin{equation}\label{eq.80}
\widehat{H}\Psi=-\frac{\hbar^{2}}{2M}\,\Delta_G\Psi+\textit{V}(\theta)\Psi=E\Psi,
\end{equation}
we obtain some second-order ordinary differential equations, just the one-dimen\-sional Schr\"odinger equations for the ``radial'' wave function $f(\theta)$. 

Let us write down explicitly these expressions. For that we must use the explicit form of the Laplace-Beltrami operators $\Delta_G$ for our spherical, pseudospherical and toroidal geometries.

On the sphere we have
\begin{eqnarray}
\Delta&=&\frac{1}{R^2}\frac{\partial^2}{\partial \theta^2}+\frac{1}{R^2}{\rm ctg}\theta\frac{\partial}{\partial\theta}+
\frac{1}{R^2\sin^2\theta}\frac{\partial^2}{\partial\phi^2}+\label{eq.81}\\
&-&\frac{2\cos\theta}{R^2\sin^2\theta}\frac{\partial^2}{\partial\phi\partial\psi}
+\frac{MR^2 \sin^2\theta+I\cos^{2}\theta}{IR^2\sin^2\theta}\frac{\partial^2}{\partial\psi^2}.\nonumber
\end{eqnarray}
In the special case of ``inertial resonance'', when $I=mR^2$, one obtains the model invariant under ${\rm SO}(3,\mathbb{R})\times {\rm SO}(3,\mathbb{R})$:
\begin{eqnarray}
\Delta_{0}&=&\frac{1}{R^2}\frac{\partial^2}{\partial\theta^2}+
\frac{1}{R^2}{\rm ctg}\theta\frac{\partial}{\partial\theta}+
\frac{1}{R^2\sin^2}\frac{\partial^2}{\partial\phi^2}+\label{eq.82}\\
&-&
\frac{2\cos\theta}{R^2\sin^2\theta}\frac{\partial^2}{\partial\phi\partial\psi}+
\frac{1}{R^2\sin^2\theta}\frac{\partial^2}{\partial\psi^2}.\nonumber
\end{eqnarray}
This is formally (and non-accidentally) identical with Laplace-Beltrami operator on ${\rm SO}(3,\mathbb{R})$, when $\left(\theta,\phi,\psi\right)$ are interpreted as Euler angles in traditional notation (respectively: nutation, precession and proper rotation).

In pseudospherical geometry we have that
\begin{eqnarray}
\Delta&=&\frac{1}{R^2}\frac{\partial^2}{\partial\theta^2}+
\frac{1}{R^2}{\rm cth}\theta\frac{\partial}{\partial\theta}+
\frac{1}{R^2{\rm sh}^2}\frac{\partial^2}{\partial\phi^2}+\label{eq.83}\\
&-&
\frac{2{\rm ch}\theta}{R^2{\rm sh}^2\theta}\frac{\partial^2}{\partial\phi\partial\psi}+
\left(\frac{M}{I}+\frac{1}{R^2}{\rm cth}^2\theta \right)\frac{\partial^2}{\partial\psi^2}.\nonumber
\end{eqnarray}
This operator is invariant under ${\rm SO}(1,2)\times {\rm SO}(2,\mathbb{R})$ (left and right respectively) if interpreted as a differential operator acting on ${\rm SO}(1,3)$. If, as already once done above, we assume the rotational kinetic energy to contribute with the negative sign, then (\ref{eq.83}) becomes
\begin{eqnarray}
\Delta^{(-)}&=&\frac{1}{R^2}\frac{\partial^2}{\partial\theta^2}+
\frac{1}{R^2}{\rm cth}\theta\frac{\partial}{\partial\theta}+
\frac{1}{R^2{\rm sh}^2}\frac{\partial^2}{\partial\phi^2}+\label{eq.84}\\
&-&
\frac{2{\rm ch}\theta}{R^2{\rm sh}^2\theta}\frac{\partial^2}{\partial\phi\partial\psi}+
\left(-\frac{M}{I}+\frac{1}{R^2}{\rm cth}^2\theta \right)\frac{\partial^2}{\partial\psi^2}.\nonumber
\end{eqnarray}

In the special case $I=mR^2$ these operators simplify respectively to
\begin{eqnarray}
\Delta_{0}&=&\frac{1}{R^2}\frac{\partial^2}{\partial\theta^2}+
\frac{1}{R^2}{\rm cth}\theta\frac{\partial}{\partial\theta}+
\frac{1}{R^2{\rm sh}^2\theta}\frac{\partial^2}{\partial\phi^2}+\label{eq.85}\\
&-&\frac{2{\rm ch}\theta}{R^2{\rm sh}^2\theta}\frac{\partial^2}{\partial\phi\partial\psi}+
\frac{1}{R^2}\frac{{\rm ch}2\theta}{{\rm sh}^2\theta}\frac{\partial^2}{\partial\psi^2},\nonumber\\
\Delta^{(-)}_{0}&=&\frac{1}{R^2}\frac{\partial^2}{\partial\theta^2}+
\frac{1}{R^2}{\rm cth}\theta\frac{\partial}{\partial\theta}+
\frac{1}{R^2{\rm sh}^2 \theta}\frac{\partial^2}{\partial\phi^2}+\label{eq.86}\\
&-&\frac{2{\rm ch}\theta}{R^2{\rm sh}^2\theta}\frac{\partial^2}{\partial\phi\partial\psi}+
\frac{1}{R^2{\rm sh}^2\theta}\frac{\partial^2}{\partial\psi^2}.\nonumber
\end{eqnarray}
The last operator, when interpreted as one acting on ${\rm SO}(1,2)$ is doubly invariant, i.e., left and right, so under ${\rm SO}(1,2)\times {\rm SO}(1,2)$. It must be stressed, it has the normal-hyperbolic signature.

For the injected torus in $\mathbb R^3$ we have the following Laplace-Beltrami operator:
\begin{eqnarray}
\Delta&=&\frac{1}{R^2}\frac{\partial^2}{\partial\theta^2}-
\frac{\sin\theta}{R(L+R\cos\theta)}\frac{\partial}{\partial\theta}
+\frac{1}{(L+R\cos\theta)^2}\frac{\partial^2}{\partial\phi^2}+\label{eq.87}\\
&-&\frac{2\sin\theta}{(L+R\cos\theta)^2}\frac{\partial^2}{\partial\phi\partial\psi}
+\left(\frac{M}{I} +\frac{\sin^2\theta}{(L+R\cos\theta)^2}\right)\frac{\partial^2}{\partial\psi^2}.\nonumber
\end{eqnarray}

Let us substitute now (\ref{eq.77}) to (\ref{eq.80}) or rather to its free version (\ref{eq.81}), (\ref{eq.83}), (\ref{eq.87}) corresponding to the spherical, pseudospherical and toroidal geometries of the ``physical'' space. In the pseudospherical case we consider also the model of hyperbolic signature (\ref{eq.84}).

Inserting (\ref{eq.77}) to (\ref{eq.81}) one obtains after a few standard manipulations the following ``radial'' equation for the function $f$:
\begin{eqnarray}
\frac{d^{2}f}{d\theta^{2}}+{\rm ctg}\theta \,\frac{df}{d\theta}-\left(\frac{m^{2}}{\sin^{2}\theta}-
\frac{2ms\cos\theta}{\sin^{2}\theta}+\left(\frac{MR^{2}}{I}+{\rm ctg}^{2}\theta\right)s^{2}\right)f+&&\label{eq.88}\\
+ \frac{2MR^{2}}{\hbar^{2}}\left(E-V(\theta)\right)f=0.&&\nonumber
\end{eqnarray}
In the resonance case $I=MR^{2}$, this becomes
\begin{equation}\label{eq.89}
\frac{d^{2}f}{d\theta^{2}}+{\rm ctg}\theta \,\frac{df}{d\theta}-\frac{m^{2}-2ms\cos\theta+s^{2}}
{\sin^{2}\theta}f+\frac{2I}{\hbar^{2}}\left(E-V(\theta)\right)f=0.
\end{equation}

Let us notice that these ordinary (one-dimensional) Schr\"{o}dinger equations are formally identical with the corresponding Schr\"{o}dinger equations for the symmetric (\ref{eq.88}) and spherical (\ref{eq.89}) three-dimensional rigid body subject to the action of $\theta$-dependent forces. In particular, for the geodetic case $V=0$, this reduces to the Schr\"{o}dinger equation for the matrix elements
\begin{equation}\label{eq.90}
D^{j}{}_{sm}=e^{is\phi}e^{im\psi}f^{j}{}_{sm}(\theta)
\end{equation}
of the $j-th$ unitary irreducible representation of ${\rm SO}(3,\mathbb{R})/{\rm SU}(2)$. These functions may be also used as a basis for some approximation methods in the non-geodetic case of some $\theta$-dependent (or even more general) potential $V$. In view of the compactness of ${\rm SU}(2)/{\rm SO}(3,\mathbb{R})$ the geodetic case is physically reasonable.

Doing the same in Lobatschevski space (pseudospherical geometry) we obtain the following one-dimensional Schr\"{o}dinger equation for $f(\theta)$:
\begin{eqnarray}
\frac{d^{2}f}{d\theta^{2}}+{\rm cth}\theta \,\frac{df}{d\theta}-\frac{m^{2}-2ms{\rm ch}\theta+\left(\pm\frac{MR^{2}}{I}+{\rm cth}^{2}\theta
\right)s^{2}}{{\rm sh}^{2}\theta}f+&&\label{eq.91}\\
+\frac{2MR^{2}}{\hbar^{2}}\left(E-V(\theta)\right)f=0.&&\nonumber
\end{eqnarray}
Let us notice that the $\pm$ signs refer respectively to the usual physical model (the $"+"$ sign) and to the strange, but geometrically more interesting model with the hyperbolic signature (the $"-"$ sign). In particular, in the resonance case $I=MR^{2}$, one obtains the following equation:
\begin{equation}\label{eq.92}
\frac{d^{2}f}{d\theta^{2}}+{\rm cth}\theta \,\frac{df}{d\theta}-\frac{m^{2}-
2ms{\rm ch}\theta+s^{2}}{sh^{2}\theta}f+
\frac{2I}{\hbar^{2}}\left(E-V(\theta)\right)f=0.
\end{equation}
The peculiarity of this model is its invariance under ${\rm SL}(2,\mathbb{R})$ as internal symmetry, just like the spherical symmetry in (\ref{eq.89}).

Let us notice that in spite of formal similarity of (\ref{eq.88}), (\ref{eq.89}) and (\ref{eq.91}), (\ref{eq.92}) (trigonometric functions) there are serious differences between spherical and hyperbolic geometry. Of course, in the Lobatschevski space any  geodetic model is purely scattering one, due to the non-compactness of pseudospherical geometry. Therefore, certainly some attractive potential is necessary if (\ref{eq.92}) is to be physically realistic.

Finally. let us substitute now the assumption (\ref{eq.77}) of cyclic variables $\phi,\psi$ to the special case (\ref{eq.87}) of physical toroidal geometry. The one-dimensional Schr\"{o}dinger equation becomes then 
\begin{eqnarray}
&&\frac{d^{2}f}{d\theta^{2}}-\frac{R\sin\theta}{L+R\cos\theta}\frac{df}{d\theta}+\label{eq.93}
\\
&&-\left(\frac{R^{2}m^{2}}{(L+R\cos\theta)^{2}}-\frac{2R^{2}ms\sin\theta}
{\left(L+R\cos\theta\right)^{2}}+R^{2}\left(\frac{M}{I}+
\frac{\left(\sin^{2}\theta\right)}
{\left(L+R\cos\theta\right)^{2}}\right)s^{2}\right)f+\nonumber\\
&&+\frac{2MR^{2}}{\hbar^{2}}(E-V(\theta))f=0.\nonumber
\end{eqnarray}
There is also some similarity to the ``resonance'' case in the spherical and hyperbolic geometry. 

It is seen that the special case $I=MR^{2}$, although it leads to certain simplification of (\ref{eq.93}) is not so special as it was in the spherical and the pseudospherical geometry (to be honest, in the pseudospherical geometry it was the situation of the ``minus sign'' in (\ref{eq.84})). In any case, this is not the case of the suddenly increasing symmetry group. This was to be expected even on the basic of purely classical arguments. Namely, as said above, the toroidal geometry is algebraic, but of the fourth degree, unlike the second degree of the spherical and Lobatschevski spaces. On the classical level even in the purely geodetic case, the simplest solutions of equations of motion were to be expressed in terms of elliptic functions.

\section*{Acknowledgements}

This paper contains results obtained within the framework of the research project
N501 049 540 financed from the Scientific Research Support Fund in 2011-2014 and the Institute of Fundamental Technological Research PAS internal project 203. The authors are greatly indebted to the Ministry of Science and Higher Education for
this financial support.

\end{document}